\documentstyle[aps,preprint]{revtex}

\begin{document}

\title{Insulator-Metal Transition in 
the One and Two-Dimensional Hubbard Models}
\author{ F.F. Assaad and    M. Imada \\
         Institute for Solid State Physics, University of Tokyo,  \\ 
         7-22-1 Roppongi, 
         Minato-ku, Tokyo 106, Japan. }

\maketitle
\begin{abstract}
We use Quantum Monte Carlo methods to determine $T=0$ Green functions,
$G(\vec{r}, \omega)$,
on lattices up to $16 \times 16$  for the 2D Hubbard model at $U/t =4$. 
For chemical potentials, $\mu$, within the Hubbard gap, 
$ |\mu | < \mu_c$,  and at {\it long} distances, $\vec{r}$,
$G(\vec{r}, \omega = \mu) \sim e^{ -|\vec{r}|/\xi_l}$ 
with critical behavior:
$\xi_l \sim | \mu - \mu_c |^{-\nu}$,  
$ \nu = 0.26 \pm 0.05$.  This result stands in agreement with the
assumption of hyperscaling with correlation exponent $\nu = 1/4$ and
dynamical exponent $z = 4$. In contrast, the generic
band insulator as well as the metal-insulator  transition 
in the 1D Hubbard model 
are characterized by $\nu = 1/2$ and 
$z = 2$. 
\end{abstract}


\newpage

At zero-temperature a continuous metal-insulator  transition 
driven by a change in chemical potential may be characterized 
by the 
compressibility, $\chi_c$, or the Drude weight, $D$.
In the Mott insulating phase both $D$ and $\chi_c$ vanish 
while they remain finite in the metallic phase \cite{Kohn,Imada}.
In order to describe the metal-insulator transition from the insulator 
side, we consider the zero-temperature 
Green function $G(\vec{r},\omega)$ \cite{Note0}. At {\it long}
distances, $|\vec{r}|$,  and for values of the  chemical
potential, $\mu$, within the charge gap, $G(\vec{r},\omega = \mu) \sim
e^{-|\vec{r}|/\xi_l}$.  The metal-insulator  transition may be
characterized by the divergence of $\xi_l$
as the critical chemical potential, $\mu_c$,
is approached from the insulating phase. 
$\xi_l$ may be interpreted as the localization length involved in 
transferring a particle over a distance $\vec{r}$ 
from the electronic system to the heat bath lying at energy $\mu$ within
the charge gap. 
Under the assumption  of hyperscaling, the above quantities are expected
to satisfy the scaling relations:
\begin{equation}
\label{Scale}
     \xi_l \sim \Delta^{-\nu}, \; \;
     \chi_c  \sim \Delta^{-\nu (z-d)}, \; \;
     D \sim \Delta^{\nu (d + z - 2)}, \; \;
\end{equation}
where $\Delta = |\mu - \mu_c|$, $d$ is the dimensionality and  $\nu$ ($z$)
corresponds to the correlation length (dynamical) exponent \cite{Imada}.  
Since the control parameter, $\Delta$, 
corresponds to the chemical potential, one  obtains the 
additional constraint $\nu z = 1$ as well as 
$\delta \sim \Delta^{\nu (d + z) -1} $, $\delta$ being the 
doping concentration. 
In this letter, based on a recently developed numerically stable
Quantum Monte Carlo (QMC) algorithm to calculate zero-temperature 
imaginary time Green functions \cite{Assaad}, 
we calculate $\xi_l$ for the two-dimensional 
repulsive Hubbard  model at $U/t = 4$.
We obtain: $\mu_c = 0.67 \pm 0.015$ in units of the hopping matrix
element and $\xi_l \sim | \mu - \mu_c |^{-\nu}$ with
$ \nu = 0.26 \pm 0.05$. On the other hand, the compressibility
data of Furukawa and Imada  
\cite{Furukawa} in the metallic phase is 
consistent with  $\chi_c \sim  |\mu - \mu_c|^{-1/2}$. 
When hyperscaling describes the transition, 
the above scaling of the compressibility also leads to $\nu = 1/4$
\cite{Imada}.
Comparison of those results puts the hyperscaling assumption on a
firmer basis. 
The present estimate of $\nu$ is a more direct determination
of the characteristic length scale. 
In contrast, the Mott transition in the one-dimensional Hubbard model 
satisfies the scaling relations (\ref{Scale}) with 
$\nu = 1/2$ and $z = 2$ \cite{Imada,Mori,Shastry,Imada1}. 
The generic
band insulator in all dimensions  equally belongs to the 
universality class $\nu = 1/2$ and $z = 2$ \cite{Imada}. 
The above results point out the anomalous  character of the Mott transition
in the two-dimensional Hubbard model. 

The Hubbard model we consider reads:
\begin{equation}
\label{HamU}
     H  - \mu N = -t \sum_{\langle\vec{i},\vec{j} \rangle,\sigma}
           c_{\vec{i},\sigma}^{\dagger} c_{\vec{j},\sigma}   +
           U \sum_{\vec{i}} 
          (n_{\vec{i},\uparrow}-\frac{1}{2}) 
          (n_{\vec{i},\downarrow} -\frac{1}{2}) 
  - \mu  \sum_{\vec{i},\sigma} c_{\vec{i},\sigma}^{\dagger} c_{\vec{i},\sigma}
\end{equation}
Here, $\langle \vec{i},\vec{j} \rangle$ denotes nearest-neighbors.
$c_{\vec{i},\sigma}^{\dagger}$ ($c_{\vec{i},\sigma}$)
creates (annihilates) an 
electron with {\it z}-component of spin $\sigma$ on site
$\vec{i}$  and $n_{\vec{i},\sigma } =  c_{\vec{i},\sigma}^{\dagger}
c_{\vec{i}\sigma}$. 
In this notation half-band filling corresponds to $\mu = 0$. 
We start by considering the  non-interacting case, $U/t = 0$. In Fourier
space, the single particle energies are given by 
$\epsilon_{\vec{k}} = -2t(\cos(\vec{k} \vec{a}_x) +
\cos(\vec{k} \vec{a}_y) ) $, $\vec{a}_x$, $\vec{a}_y$ being
the lattice constants. In this letter, the length scale is
set by: $ |\vec{a}_x| = |\vec{a}_y | = 1$.   At zero temperature, 
an insulator-metal transition will occur when $\mu \rightarrow \mu_c =
4t$. For those chemical potentials, the  zero-temperature Green function 
\cite{Note0}
at $\omega = \mu$ is given by:
\begin{equation}
\label{xiU0}
  G(\vec{r}, \omega = \mu ) = 
             \frac{2}{N} \sum_{\vec{k}} 
       \frac{ e^{ i \vec{k}  \vec{r} } }
            { \epsilon_{\vec{k}} - \mu }
\end{equation}
where $N$ denotes the number of sites of the square lattice and
the factor $2$ corresponds to the summation over the spin degrees of
freedom.  
Numerically, one obtains:
$G(\vec{r}, \omega = \mu ) \sim  e^{-|\vec{r}|/\xi_l}$ 
with critical behavior: $\xi_l \sim | \mu - \mu_c |^{-1/2}$ \cite{Note1}.
One may check that this example of
a band insulator-metal transition satisfies the above scaling
relations (\ref{Scale}) and is characterized by the exponent $\nu=1/2$.
At finite values of $U/t$ and half-band filling the 
antiferromagnetic Hartree-Fock approximation 
equally yields the
exponent $\nu = 1/2$ for $ \xi_l$. 
However, this approximation does not
satisfy the hyperscaling assumption. 

The physical interpretation of $\xi_l$ 
is facilitated by considering the single impurity 
Hamiltonian (Fano-Anderson model) \cite{Anderson,Mahan}:
\begin{equation}
       H = \sum_{\vec{k}} \epsilon_{k} 
            c_{\vec{k}}^{\dagger} c_{\vec{k}} 
           + \epsilon_b b^{\dagger} b    + 
           \frac{t_b}{\sqrt{N} }
           \sum_{\vec{k}} 
\left( c_{\vec{k}}^{\dagger} b  + b^{\dagger} c_{\vec{k}} \right).
\end{equation}
Here, $b^{\dagger}$ creates an electron in the impurity state at the 
origin and  energy $\epsilon_b$.
The hybridization between the
localized state and the band electrons alters the 
energy of the impurity level to the value:
$ E_b = \epsilon_b + \frac{t_b^2}{N}\sum_{\vec{k}} 
\frac{1}{E_b - \epsilon_{\vec{k}} } $ . We will assume 
$ E_b > \epsilon_{\vec{k}}$  for all  $\vec{k} $.
When all single particle states of the valence band are filled
and the impurity single particle state empty, the probability amplitude
of transferring a band electron at site $\vec{r}$
to the impurity state is:
\begin{equation}
\label{xiFA}
\frac{\langle \Psi_0 | b^{\dagger} c_{\vec{r}} |\Psi_0 \rangle }
     {\langle \Psi_0 | \Psi_0 \rangle }   =
        -\alpha(E_b) \frac{t_b}{N} \sum_{\vec{k}} 
       \frac{ e^{ i \vec{k}  \vec{r} } }
            { \epsilon_{\vec{k}} - E_b }
\end{equation}
where the normalization factor is given by 
$\alpha^{-1}(E_b) = 1 +  \frac{t_b^2}{N}\sum_{\vec{k}}
\frac{1}{ (E_b - \epsilon_{\vec{k}})^2 }$.  
Comparison between equations (\ref{xiFA}) and (\ref{xiU0})
show that the spatial dependence of the two quantities is
identical.  
$\xi_l$ may thus be interpreted as the localization length 
involved in transferring a particle over a distance $\vec{r}$ 
from the valence band to the 
heat bath (see equation (\ref{xiU0})), or to the impurity state
(see equation (\ref{xiFA})). This definition of $\xi_l$ bears
some similarity to that applied in finite size scaling studies of
Anderson localized states \cite{Herbert,MacKinnon}. 
When the imaginary part of the Green function vanishes in
the insulating phase, $\xi_l$ may be used to study the 
metal-insulator transition. 

To obtain an estimate of the critical exponent $\nu$ for the Hubbard model,
we require an accurate determination of the critical chemical 
potential, $\mu_c$ as well as  of the localization length, $\xi_l$.  
Both quantities may be obtained from the knowledge of:
\begin{equation}
\label{Grtau}
G_{\sigma}(\vec{r}, \tau)  =  \Theta(\tau)
   \frac{ \langle \Psi_0 |c_{\vec{r},\sigma}(\tau) 
                          c_{\vec{0},\sigma}^{\dagger} |  \Psi_0 \rangle}
        { \langle \Psi_0 |  \Psi_0 \rangle}
         - \Theta(-\tau)
   \frac{ \langle \Psi_0 | c_{-\vec{r},\sigma}^{\dagger}(-\tau) 
                           c_{ \vec{0},\sigma} |  \Psi_0 \rangle}
        { \langle \Psi_0 |  \Psi_0 \rangle },
\end{equation}
where $c_{\vec{r},\sigma}(\tau)  = 
e^{ \tau H } c_{\vec{r},\sigma} e^{-\tau H }$.
Here $ |\Psi_0 \rangle $ denotes the ground state of the half-filled
($\mu = 0$) Hubbard Hamiltonian (\ref{HamU}).  
The above quantity may be efficiently calculated 
with QMC methods. 
Since we are at half-band filling, the sign problem does
not occur and we are able to consider lattice sizes up to 
linear dimension $L=16$, namely $N = 16 \times 16$ without
any serious difficulties. 
The calculation of imaginary time Green function 
in  the zero-temperature auxiliary field QMC algorithm \cite{Sandro,Koonin} 
was
first reported by Deisz et al. \cite{John}.
However,
their approach does not incorporate a numerical stabilization
scheme and they are thus restricted to small values of $\tau$ (i.e.
$\tau t  \sim 2.5$ ). The authors have developed a numerically stable
QMC algorithm for the calculation of $G_{\sigma}(\vec{r}, \tau)$. 
The details of the algorithm may be found in reference \cite{Assaad}.
All our calculations were performed with periodic boundary conditions.

From the knowledge of $G_{\sigma}(\vec{r}, \tau)$ on an $N$-site 
lattice, we may obtain an estimate of the charge gap. 
We denote by $| \Psi_{n}^{N} \rangle$ the eigenvector of the 
Hamiltonian $H$ with eigenvalue $E_n^{N}$ in the $N$-particle Hilbert space.
With this notation, 
\begin{equation}
  G(\vec{r} = 0, \tau) \equiv \sum_{\sigma} G_{\sigma}(\vec{r}=0, \tau) =
     \frac{1}{N} \sum_{n,\vec{i},\sigma}
 | \langle \Psi_{0}^{N} | c_{\vec{i},\sigma} | \Psi_{n}^{N +1} \rangle |^2
   \exp \left( -\tau \left( E_n^{N+1} - E_0^N \right) \right).
\end{equation}
for  $\tau >0$.
Fig. 1a, plots $G(\vec{r} = 0, \tau)$ for a $16 \times 16$ lattice at
$U/t = 4$.  
We may obtain a reliable estimate of the charge gap for this
lattice size
by fitting $G(\vec{r} = 0, \tau)$ 
to the form $ e^{-\Delta_c \tau} $ with 
$\Delta_c \equiv E_0^{N+1} - E_0^{N}$ for {\it large} values of $\tau$.
Fig. 1b shows $ \Delta_c$ as a
function of linear lattice size. Extrapolation to the thermodynamic limit
yields: $\Delta_c/t = 0.67 \pm 0.015$.
This result stands in good agreement with
the quoted result of Furukawa and Imada \cite{Furukawa}: 
$ \Delta_c/t = 0.58 \pm 0.08$.
Since in the notation of equation (\ref{HamU}) the Hubbard model 
satisfies particle-hole symmetry at $\mu = 0$, 
the critical chemical potential is nothing but the charge gap:
$\mu_c \equiv \Delta_c$. 

For values of the chemical potential within the charge gap,
$ |\mu | < \mu_c $,  the zero temperature Green function is real and may be 
obtained through the relation:
\begin{equation}
\label{Inter}
     G(\vec{r}, \omega = \mu) = -\int_{-\infty}^{\infty} {\rm d}\tau
     G(\vec{r}, \tau) e^{\tau \mu}.
\end{equation}
The Green function $G(\vec{r}, \tau)$
is computed at half-band filling where the sign problem is not present
and the statistical uncertainty does not grow exponentially with lattice
size. However, since we are multiplying the QMC data by the factor
$e^{\tau \mu}$, the statistical uncertainty will grow exponentially with
increasing values of $\tau$ for $\tau \mu > 0$. 
For each lattice size $L$,
we have considered the largest distance $\vec{r} = (L/2,L/2)$.
For this distance,
$G(\vec{r} = (L/2,L/2), \tau )$ is plotted in Fig. 2.  Due to
particle-hole symmetry at $\mu =0$, $G(\vec{r} = (L/2,L/2), \tau ) = $
$-G(\vec{r} = (L/2,L/2), -\tau ) $. For the imaginary time integration, 
(see equation (\ref{Inter}))  and values of the chemical potential
$ |\mu| < 0.65 t$, a cutoff of $ \tau t = 10$ proved to be sufficient
for the determination of the Green function (see Fig. 2).
$G(\vec{r}= (L/2,L/2), \omega = \mu)$ as a function of lattice size
is plotted in Fig. 3  for several values of $\mu$. 
For lattice sizes ranging from
$L = 6$ to $L = 16$,  an exponential decay
may be seen within the quoted statistical uncertainty. 
From this data, we obtain an estimate of the localization
length $\xi_l$. 
With the above determined value of $\mu_c$, we plot
$| \mu - \mu_c | $ versus $\xi_l$ (see Fig. 4) on a log-log plot. 
For all considered chemical potentials, $\xi_l/a < 8\sqrt{2} $
which corresponds to our largest considered distance ($L=16$).
The slopes in Fig. 4  correspond to values of the critical exponent 
$\nu=1/2$ and $\nu=1/4$. 
The QMC data is consistent with
$\nu = 1/4$ and  seems to rule out the possibility $\nu=1/2$.
A statistical analysis yields: $\nu = 0.26 \pm 0.05$.

The choice  $\vec{r} = (L/2,L/2)$ is very convenient since apart from a
sign change between $L = 4n +2$ and $L = 4n$ lattices the
exponential decay of
$G(\vec{r} = (L/2,L/2) ,\omega = \mu)$  as a function of $L$,
is not masked by a non-trivial oscillation.
For other choices of $\vec{r}$ we expect to obtain the same results 
since due to the point group symmetry of the square lattice, all 
$\vec{k}$-points in the Brillouin zone contribute to 
$G(\vec{r} = (L/2,L/2), \tau )$. We could however not check this point
numerically.

To confirm the validity of our approach, we consider the 
Mott transition in the one-dimensional 
Hubbard model at $U/t = 4$. 
A similar QMC calculation as described above
for the two-dimensional case yields a value of the correlation 
length exponent consistent with  $\nu = 1/2$ (see Fig. 5). 
We have obtained $\mu_c/t = 0.66 \pm 0.015$ which is consistent with the
exact result of Lieb and Wu \cite{LiebWu}: $\mu_c/t = 0.643$. 
Chains of linear length up
to $L = 24$ were considered.
This result stands in agreement with the assumption 
of hyperscaling with exponents
$\nu = 1/2$ and $z = 2$ \cite{Imada,Mori,Shastry,Imada1}. 

In conclusion, we have determined the correlation length exponent 
from the knowledge of $\xi_l$  in the 
insulating phase of the two-dimensional
Hubbard model and obtained:  $\nu = 0.26 \pm 0.05$. 
Under the assumption of hyperscaling 
with exponents $\nu = 1/4$ and $z = 4$, this  result stands in agreement
with compressibility measurements in the metallic phase \cite{Furukawa}.
We have shown that a similar calculation for the one-dimensional 
Hubbard model, yields results consistent with $\nu = 1/2$. 
Several anomalous aspects of the metal-insulator transition
are inferred when it is 
characterized by the universality class, $\nu = 1/4$ and $z = 4$
\cite{Imada}.
Based on a single-particle theory, 
the exponents $\nu = 1/4$, $z = 4$ are consistent with the interpretation 
of the Mott transition driven
by the divergence of the effective mass as opposed to the  vanishing
of the number of charge carriers. This statement is supported by
the compressibility in the metallic phase \cite{Furukawa} 
as well as by the high frequency Hall coefficient \cite{Assaad1}.
Another consequence of the exponent $\nu = 1/4$, is the behavior
of the Drude weight in the vicinity of the Mott transition: 
$D \sim \delta^2$,   $\delta$ being the hole-density. 
As a by-product, we have produced an accurate estimate of the
charge gap for the two-dimensional Hubbard model at $U/t =4$:
$\Delta_c/t =  0.67 \pm 0.015$. 
From the technical point of view, we have introduced an efficient method
to obtain information on the nature of the metal-insulator transition
by approaching the transition from the insulator side.
The most important fact, is that for models which show particle-hole
symmetry, such as dimerized Hubbard models, the here presented method
is not plagued by the sign problem and large lattice sizes may be
considered. 

F.F.A. thanks the JSPS for financial support.
The numerical calculations were carried out on the Fujitsu VPP500 of the
Supercomputer Center of the Institute for Solid State Physics, Univ.
of Tokyo. This work is supported by a Grant-in-Aid for Scientific
Research on the Priority Area `Anomalous Metallic State near the Mott
Transition' from the Ministry of Education, Science and Culture, Japan.

\subsubsection*{Figure captions}
\newcounter{bean}
\begin{list}%
{Fig. \arabic{bean}}{\usecounter{bean}
                   \setlength{\rightmargin}{\leftmargin}}

\item  a) $\ln G(\vec{r} =0, \tau)$  as a function of $\tau$ for 
the half-filled ($\mu = 0$) 2D Hubbard model at $U/t =4$ on a 
$16 \times 16$  lattice. 
The solid line corresponds to a least square fit of $G(\vec{r} =0, \tau)$ 
to the form $\exp(-\Delta_c \tau)$ at {\it large} values of $\tau$. 
b) $\Delta_c$  as a function of linear lattice size $L$.
The solid circle  at $1/L = 0$ 
corresponds to $\Delta_c$ as obtained in reference \cite{Furukawa}. 

\item  
$ | G(\vec{r} =(L/2,L/2),\tau) |$ as a function of system size and imaginary
time $\tau$ for the 2D Hubbard model.

\item 
$ \ln | G(\vec{r} =(L/2,L/2), \omega = \mu) |$ as a function of distance 
and chemical potential for the 2D Hubbard model.
The solid lines correspond
to a least square fit of $| G(\vec{r} =(L/2,L/2), \omega = \mu) |$ 
to the form $ \exp \left( - |\vec{r}|/\xi_l \right) $ for $L>4$.

\item  Localization length $\xi_l$ 
versus $ | \mu - \mu_c| $ for the 2D Hubbard model.
The solid lines correspond to two values of the correlation 
length exponent: $\nu = 1/4$ and $\nu = 1/2$.
The solid circles correspond to the QMC data. 

\item  Same as Fig. 4 but for the 1D  Hubbard model.

\end{list}


\begin{thebibliography}{99}

\bibitem{Kohn}
W. Kohn, Phys. Rev. {\bf 133A}, 171, (1964). 

\bibitem{Imada}
M. Imada,  J. Phys. Soc. of Jpn. {\bf 64}, 2954 (1995).
 
\bibitem{Note0} We define the zero-temperature Green function by:
$ i G(\vec{r}, \omega ) $  $ = \sum_{\sigma} \int {\rm d}t  e^{i \omega t}$ 
$ \langle \Psi_0 | T c_{\sigma,\vec{r}}(t)  
c_{\sigma,\vec{0}}^\dagger(0) | \Psi_0  \rangle $.
Here, the notation is standard \cite{Mahan} and 
$ | \Psi_0 \rangle $ corresponds to the ground state in the insulating phase. 
                
\bibitem{Assaad} 
F.F. Assaad and M. Imada, to appear in J. Phys. Soc.  Jpn. 
(cond-mat/9508113)

\bibitem{Furukawa} 
N. Furukawa and M. Imada, J. Phys. Soc. Jpn. {\bf 62}, 2557, (1993).

\bibitem{Mori} M. Mori, H. Fukuyama and M. Imada, J. Phys. Soc. of Jpn.
{\bf 63}, 1639, (1994).

\bibitem{Shastry} B.S. Shastry and B. Sutherland, Phys. Rev. Lett. 
{\bf 65}, 243, (1990).

\bibitem{Imada1}  M. Imada, J. Phys. Soc. Jpn. {\bf 63}, 3059, (1994).

\bibitem{Note1}
For the dispersion relation $\epsilon_{\vec{k}} = \frac{\vec{k}^2}{2m} $ the 
integration may be done analytically to obtain the result
$\xi_l \sim  | \mu - \mu_c |^{-1/2}$, with $\mu_c = 0$. 

\bibitem{Anderson}
P.W. Anderson, Phys. Rev. {\bf 124}, 41, (1961). 

\bibitem{Mahan} G.D. Mahan, Many-Particle Physics, Plenum Press, New York, 
(1981).

\bibitem{Herbert} D.C. Herbert and R. Jones, J. Phys. C {\bf 4}, 1145, (1971).

\bibitem{MacKinnon}  A. MacKinnon and B. Kramer, Phys. Rev. Lett.
{\bf 21}, 1546, (1981).
 
\bibitem{Sandro}
S. Sorella, S. Baroni, R. Car, and M. Parrinello,
Europhys. Lett. {\bf 8} (1989) 663.
S. Sorella, E. Tosatti, S. Baroni, R.
Car, and M. Parinello, Int. J. Mod. Phys. B{\bf 1} (1989) 993.

\bibitem{Koonin}
G. Sugiyama and S.E. Koonin, Anals of Phys. {\bf 168} (1986) 1.

\bibitem{John} J.J. Deisz, W. von der Linden, R. Preuss and W. Hanke,
to appear in {\it Computer simulations in Condensed Matter Physics VIII},
Eds. D.P. Landau, K.K. Mon and H.B. Sch\"uttler
(Spinger Verlag, Heidelberg, Berlin, 1995). 

\bibitem{LiebWu}  E.H. Lieb and F.Y. Wu, Phys. Rev. Lett. {\bf 20},
1445, (1968). 

\bibitem{Assaad1} 
 F.F. Assaad and M. Imada, Phys. Rev. Lett. {\bf 74}, 3872, (1995).

\end{thebibliography}
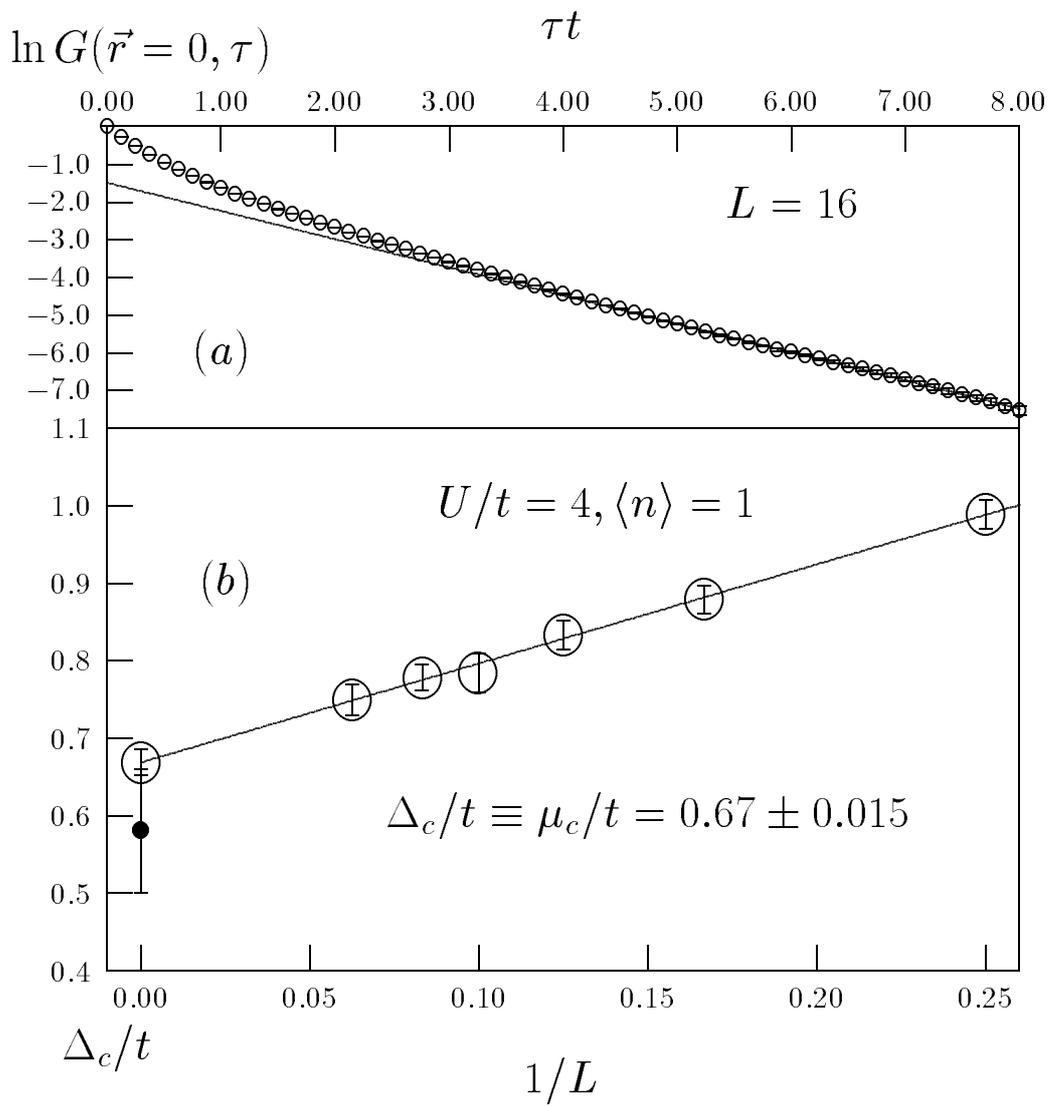
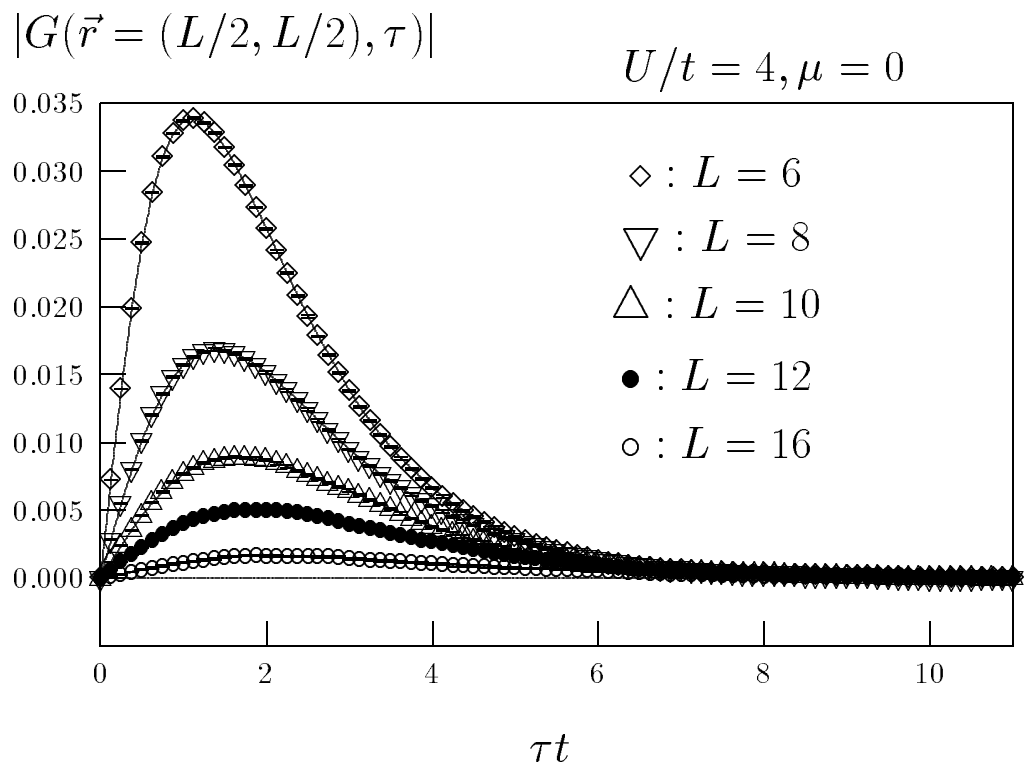
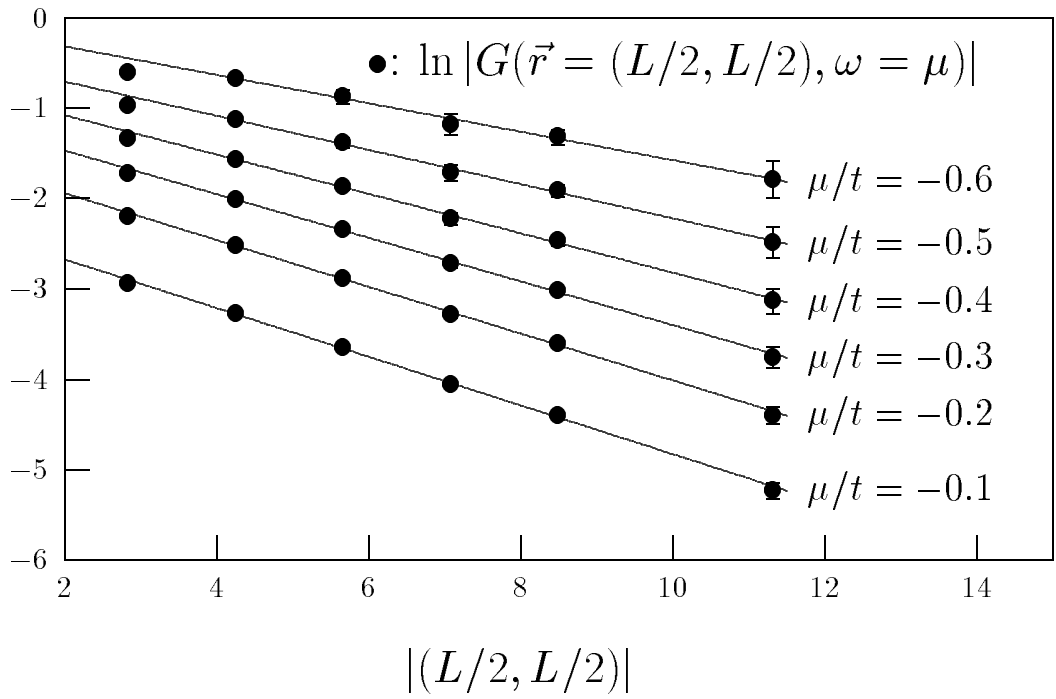
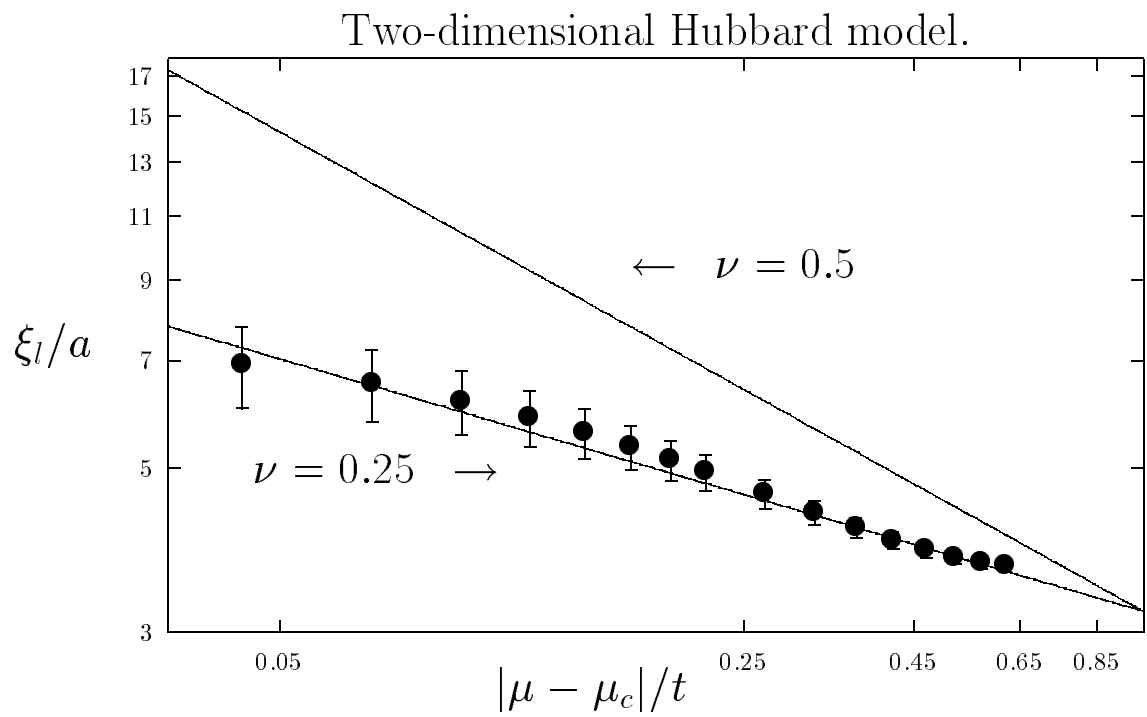
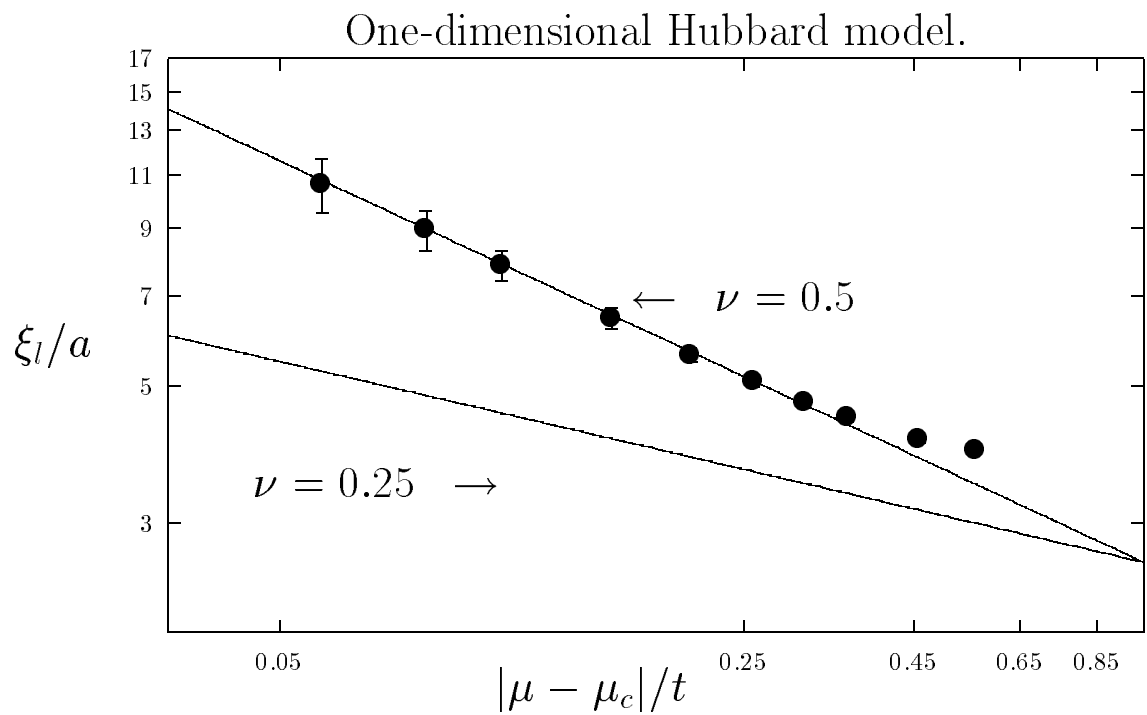
\end{document}